\documentclass[prl,aps,twocolumn]{revtex4-2}

\usepackage{graphicx}
\usepackage{xcolor}
\usepackage{caption}
\usepackage{subcaption}
\begin{document}
\title{Long-range interactions in a quantum gas mediated by diffracted light \\
}
\author{G.R.M. Robb, J.G.M. Walker, G.-L. Oppo \& T. A. Ackemann}
\affiliation{SUPA and Department of Physics, University of Strathclyde,Glasgow G4 0NG, Scotland, UK}

\begin{abstract}
A BEC interacting with an optical field via a feedback mirror can be a realisation of the quantum Hamiltonian Mean Field (HMF) model, a paradigmatic model of long-range interactions in quantum systems. We demonstrate that the self-structuring instability displayed by an initially uniform BEC can evolve as predicted by the quantum HMF model, displaying quasiperiodic "chevron" dynamics for strong driving. For weakly driven self-structuring, the BEC and optical field behave as a two-state quantum system, regularly oscillating between a spatially uniform state and a spatially periodic state. It also predicts the width of stable optomechanical droplets and the dependence of droplet width on optical pump intensity. The results presented suggest that optical diffraction-mediated interactions between atoms in a BEC may be a route to experimental realisation of quantum HMF dynamics and a useful analogue for studying quantum systems involving long-range interactions.
\end{abstract}


\date{\today}

\maketitle

Systems involving long-range interactions, such as those occurring in gravitational physics or plasma physics, display several unusual behaviours e.g. extremely slow relaxation and existence of quasi-steady states \cite{Dauxois2002}. Recently, there has been significant interest in quantum systems involving long range interactions e.g. ion chains, Rydberg gases and cold atomic gases enclosed in optical cavities \cite{Defenu2021}. 

The Hamiltonian Mean Field (HMF) model \cite{Dauxois2002} was introduced as a generic classical model of long-range interacting systems e.g. self-gravitating systems \cite{Chavanis2005}. It involves $N$ particles on a ring which experience a pairwise cosine interaction. It also arises as a model of a system of X-Y rotors coupled with infinite range. Extension of the HMF model to describe quantum systems was first carried out by Chavanis \cite{Chavanis2011_1,Chavanis2011_2} and the dynamics of this quantum HMF model was investigated more recently by Plestid \& O'Dell \cite{Plestid2018,Plestid2019} who demonstrated that the model exhibited violent relaxation of an initially homogeneous state to a structured state and possessed bright soliton solutions. 

Cold atomic gases enclosed in cavities exhibit phenomena demonstrating universal behaviours common to many different physical systems e.g. the behaviour of a cold, thermal gas in a cavity undergoing viscous momentum damping induced by optical molasses beams is related to the Kuramoto model \cite{Robb2004,vonCube2004} which describes synchronisation of globally coupled phase oscillators. It has been shown \cite{Bachelard2010,Schutz2014} that in the absence of momentum damping, a thermal gas in a cavity can exhibit dynamics similar to that of the classical HMF model. In the case of a quantum degenerate gas, e.g. a Bose-Einstein Condensate (BEC), its dynamical behaviour in a cavity has been mapped onto the Dicke-model describing coupled spins and superradiance \cite{Baumann2010}, but to date no experimental realisation of the quantum HMF model has been described or proposed. 

\begin{figure}
\centering
\includegraphics[width=0.7\columnwidth]{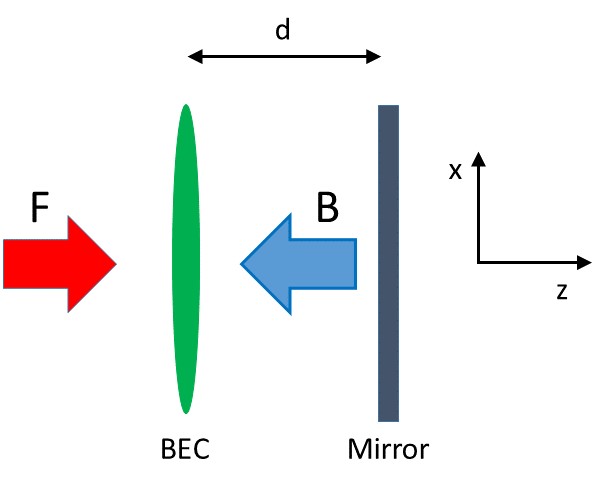}
\caption{Schematic diagram of the single mirror feedback (SMF) configuration showing a BEC interacting with a forward propagating optical field ($F$) and a retroreflected/backward propagating optical field ($B$).}
\label{fig:schematic}
\end{figure}

Here we investigate a system consisting of a BEC interacting with an optical field via single mirror feedback (SMF) as shown schematically in Fig.~\ref{fig:schematic}. In this BEC-SMF system, coupling between atoms arises due to diffraction, involves many transverse modes and optical forces directed perpendicular to the propagation direction of the optical fields. This is significantly different from cavity systems (such as e.g. \cite{vonCube2004,Baumann2010}), where the dominant coupling between atoms arises from interference between a pump field and cavity modes. We show that under certain conditions, the equations describing the dynamics of the BEC and the optical fields can be mapped onto the quantum HMF model \cite{Chavanis2011_1,Chavanis2011_2,Plestid2018}. Using this connection, we then investigate dynamical instabilities of initially homogeneous distributions of BEC density and optical intensity and also the existence of spatially localised states reminiscent of quantum droplets observed in dipolar BECs \cite{2016Ferrier,2016Chomaz}. 
The model we use to describe the BEC-SMF system was originally studied in \cite{Robb2015} as an extension of that used to study self-structuring of a classical, thermal gas, observed experimentally in \cite{Labeyrie2014}, with the thermal gas replaced with a BEC. 
We consider a BEC with negligible atomic collisions and describe  the evolution of the BEC wavefunction, $\Psi(x,t)$ with the Schrodinger equation :
\begin{equation}
i \hbar \frac{\partial \Psi(x,t)}{\partial t} = - \frac{\hbar^2}{2 m} \frac{\partial^2 \Psi(x,t)}{\partial x^2} + \frac{\hbar \delta}{2} s(x,t) \Psi(x,t)
\label{eqn:sch}
\end{equation}
where $m$ is the atomic mass, $\delta= \omega - \omega_a$ is detuning, $s(x,t) = |F|^2 + |B(x,t)|^2$ is the atomic saturation parameter due to the forward and backward optical fields where $|F|^2 = \frac{I_F}{I_{sat} \Delta^2}$ , $|B|^2 = \frac{I_B}{I_{sat} \Delta^2}$ and $I_F$, $I_B$ are the intensities of the forward (F) and backward (B) fields respectively. $I_{sat}$ is the saturation intensity on resonance, $\Delta = \frac{2 \delta}{\Gamma}$ and $\Gamma$ is the decay rate of the atomic transition. 
It has been assumed that  $|\Delta| \gg 1$  and that  consequently $s \ll 1$  so  that  the atoms remain in their ground state.  In addition, longitudinal  grating  effects  due  to  interference  between  the counterpropagating optical fields are neglected.

In order to describe the optical field in the gas we assume that the gas is sufficiently thin that diffraction can be neglected, so that the forward field transmitted through the cloud is
\begin{equation}
F_{tr}= \sqrt{p_0} \exp \left(-i \chi_0 n(x,t) \right)
\label{eqn:Ftr}
\end{equation}
where $p_0 = |F(z=0)|^2$ is the scaled pump intensity, $\chi_0 = \frac{b_0}{2 \Delta}$ is the susceptibility of the BEC, $b_0$ is the optical thickness of the BEC at resonance and $n(x,t)= |\Psi(x,t)|^2$ is the local BEC density, which for a BEC of uniform density is $n(x,t) = 1$.

The backward field, $B$, at the BEC completes the feedback loop. As the field propagates a distance $2d$ from the BEC to the  mirror  and  back,  optical  diffraction plays a critical role by converting phase modulations to amplitude modulations and consequently optical dipole forces.  The relation between the Fourier components of the forward and backward fields at the BEC is 
\begin{equation}
B(q) = \sqrt{R} F_{tr}(q) e^{i \frac{q^2 d}{k_0}}
\label{eqn:B}
\end{equation} 
where $R$ is the mirror reflectivity, $q$ is the transverse wavenumber and $k_0 = \frac{2 \pi}{\lambda_0}$.
It was shown in \cite{Robb2015} that this system exhibits a self-structuring instability where the optical fields and BEC density develop modulations with a spatial period of $\Lambda_c = \frac{2 \pi}{q_c}$, where the critcal wavenumber, $q_c$, is
\begin{equation}
q_c = \sqrt{\frac{\pi}{2} \frac{k_0}{d}}.
\label{eqn:qc}
\end{equation}
The reason for this instability is that BEC density modulations (which produce refractive index modulations) with spatial frequency $q_c$, produce phase modulations in $F_{tr}$ which are in turn converted into intensity modulations of $B$ (see Eq.~(\ref{eqn:B})). These intensity modulations produce dipole forces which reinforce the density modulations, resulting in positive feedback and instability of the initial, homogeneous state. 
A condition of this instability is that the pump intensity exceeds a threshold value, $p_{th}$ \cite{Robb2015}, which for $q=q_c$ can be written as 
\begin{equation}p_{th} = \frac{2 \omega_r }{b_0 R \Gamma} ,
\end{equation}
where $\omega_r=\frac{\hbar q_c^2}{2m}$.

The optomechanical self-structuring exhibited by the BEC-SMF model of Eq.~(\ref{eqn:sch})-(\ref{eqn:B}) derived in \cite{Robb2015} can be reduced to that of the quantum HMF model, originally proposed in \cite{Chavanis2011_1,Chavanis2011_2} and revisited in \cite{Plestid2018,Plestid2019}. We express the optical intensity, $s(x,t)$, in terms of $n$ (density) using Eq.~(\ref{eqn:Ftr}). Assuming $\chi_0 n \ll 1$ as in \cite{Zhang2018}, then 
$
F_{tr} \approx \sqrt{p_0} (1 + i \chi_0 n(x,t)) .
$
It is assumed that the BEC density and (backward) optical field consist of a spatially uniform component and a spatial modulation with spatial frequency, $q_c$, so
\begin{equation}
B(q_c) = \sqrt{R} F_{tr}(q_c) e^{i \frac{q_c^2 d }{k_0}} = i \sqrt{R} F_{tr}(q_c)
\label{eqn:F_to_B}
\end{equation}
i.e. phase modulation of $F_{tr}$ becomes amplitude modulation of $B$.
Expressing 
\begin{eqnarray*}
F_{tr}(x,t) = F_{tr}^{(0)} 
+ F_{tr}^{(q_c)} e^{i q_c x} +
F_{tr}^{(-q_c)} e^{-i q_c x} \\
n(x,t) = 1 
+ n^{(q_c)} e^{i q_c x} +
{n^{(q_c)}}^* e^{-i q_c x}
\end{eqnarray*}
then substitution of the above into Eq.~(\ref{eqn:Ftr}) shows that
\begin{equation}
\left. 
\begin{array}{c}
F_{tr}^{(0)} = \sqrt{p_0} (1 + i \chi_0) \approx \sqrt{p_0} \\
F_{tr}^{(q_c)} = i \sqrt{p_0} \chi_0 n^{(q_c)} \\
F_{tr}^{(-q_c)} = i \sqrt{p_0} \chi_0 {n^{(q_c)}}^* 
\end{array}
\right \rbrace.
\label{eqn:nq_to_Fq}
\end{equation}
Using a similar expansion of $B(x,t)$ and then Eq.~(\ref{eqn:F_to_B}),(\ref{eqn:nq_to_Fq}) produces
\[
B = \sqrt{R p_0} 
- \sqrt{R p_0} \chi_0 n^{(q_c)} e^{i q_c x}
- \sqrt{R p_0} \chi_0 {n^{(q_c)}}^* e^{-i q_c x} .
\]
Writing $n^{(q_c)} = |n^{(q_c)}|e^{-i \phi}$, then
\begin{equation}
B = \sqrt{Rp_0} - 2 \sqrt{R p_0} \chi_0 |n^{(q_c)}| \cos (q_c x - \phi) ,
\label{eqn:n_to_B}
\end{equation}
which allows us the optical field intensities in Eq.~(\ref{eqn:sch}) to be written in terms of the BEC density :
\begin{equation}
s(x,t) \approx p_0 + R p_0 - 4 R p_0 \chi_0 |n^{(q_c)}| \cos(q_c x - \phi).
\label{eq:pot}
\end{equation}
Note that if the assumption $\chi_0 n \ll 1$ was relaxed, additional terms with spatial frequency $2 q_c$ would also be present. As $n^{(q_c)}$ is described by 
$
n^{(q_c)} = \frac{1}{L} \int_0^{L} |\Psi(x,t)|^2  e^{-i q_c x} \; dx
$
, where $L$ is the BEC length then
\[
|n^{(q_c)}| \cos(q_c x - \phi) = \frac{1}{2 \pi}  \int_0^{2 \pi} |\Psi(\theta',t)|^2  \cos (\theta - \theta')) \; d \theta'
\]
where $\theta = q_c x$ and it has been assumed that $\Psi$ is spatially periodic with period $\Lambda_c$. Consequently, Eq.~(\ref{eq:pot}) can be written as 
\begin{equation}
\label{eqn:s_to_Phi}
s(x,t) = p_0 + R p_0 - 4Rp_0 \chi_0 \Phi(\theta,t)
\end{equation}
where the non-local potential , $\Phi(\theta,t)$, is
$
\Phi(\theta,t) = \frac{1}{2 \pi}  \int_0^{2 \pi} |\Psi(\theta',t)|^2  \cos (\theta - \theta')) \; d \theta' .
$
The constant term in Eq.~(\ref{eqn:s_to_Phi}) results in a constant potential energy contribution to Eq.(\ref{eqn:sch}), which can be eliminated by the transformation, $\Psi$ via $\Psi = \Psi' \exp (-i \frac{(1+R) p_0 \delta}{2} t)$, so that the Schrodinger equation from Eq.~(\ref{eqn:sch}) becomes a Gross-Pitaevskii Equation (GPE) analogue
\begin{equation}
i \frac{\partial \Psi'}{\partial t} = 
- \omega_r \frac{\partial^2 \Psi'}{\partial \theta^2} - \epsilon \Phi(\theta,t) \Psi' 
\label{eqn:GPE}
\end{equation}
where
$
\epsilon = 2 \delta R p_0 \chi_0 = \frac{R p_0 b_0 \Gamma}{2} = \frac{p_0}{p_{th}} \omega_r .
$
Eq.(\ref{eqn:GPE}) has the same effective GPE-like form as that of the quantum HMF model \cite{Chavanis2011_1,Plestid2018}. Note that $\epsilon > 0$ always, which corresponds to the case of the ferromagnetic quantum HMF model.
The order parameter or magnetization, $M$, is essentially the Fourier component of the BEC density with spatial frequency $q_c$ \cite{Chavanis2011_2,Plestid2018},
\begin{equation}
M = \left| \frac{1}{2 \pi} \int_0^{2 \pi} |\Psi|^2 e^{i \theta} \; d \theta \right| .
\label{eqn:M_q}
\end{equation}

\begin{figure}
\begin{center}
\includegraphics[width=\columnwidth]{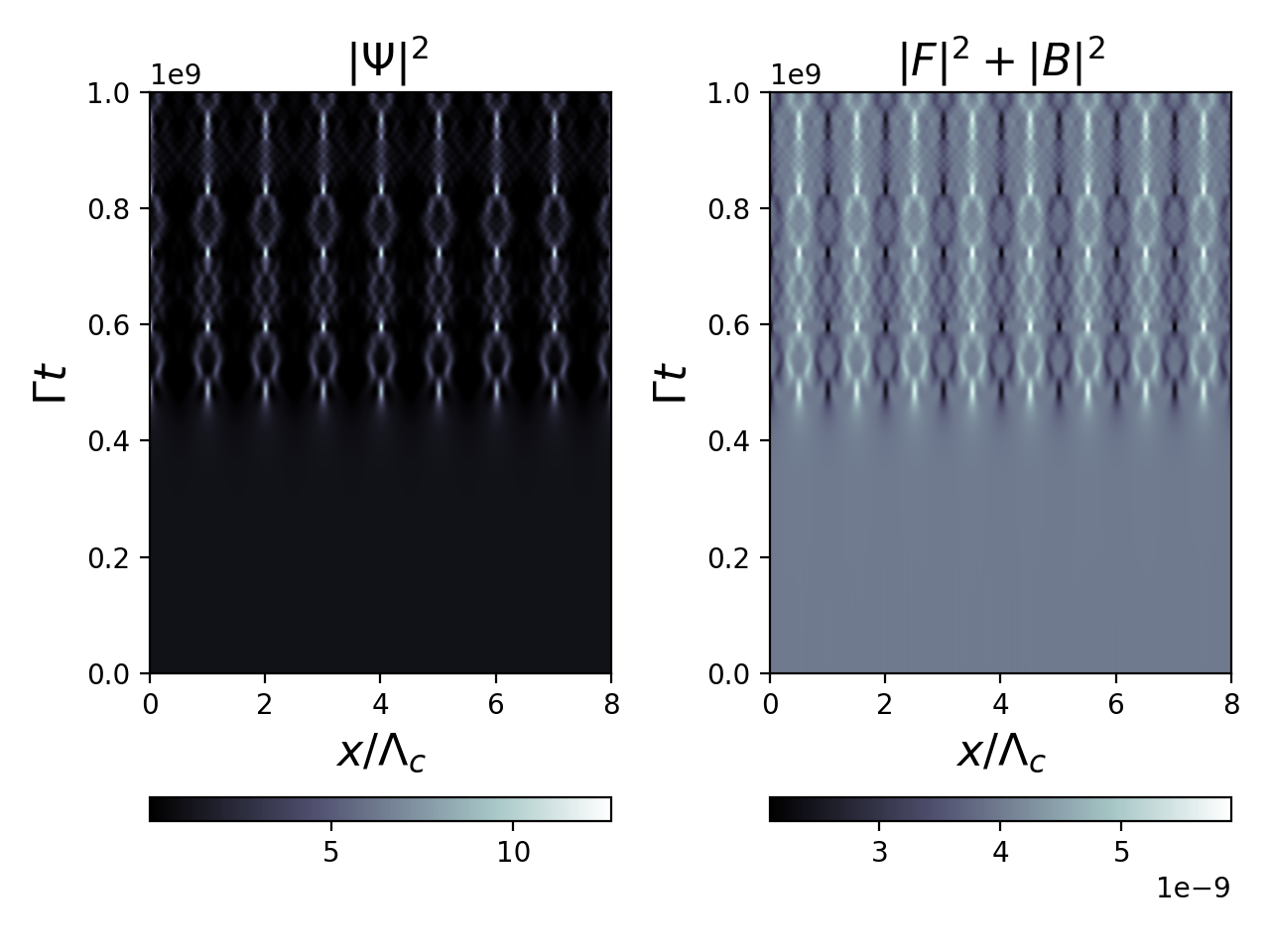}
\caption{Evolution of BEC density and optical intensity for strong driving, calculated from Eq.~(\ref{eqn:sch})-(\ref{eqn:B}). Parameters used : $b_0 = 100$, $\Delta = 500$, $p_0 = 10 p_{th} = 2 \times 10^{-9}$, $R = 1$, $\frac{\omega_r}{\Gamma}= 10^{-8}$.}
\label{fig:intens_and_dens_BEC_strong}
\end{center}
\end{figure}

In order to demonstrate that Eq.~(\ref{eqn:sch})-(\ref{eqn:B}) can exhibit dynamical behaviour associated with the quantum HMF model, we consider two example cases :  strong driving, far above threshold i.e. $p_0 \gg p_{th}$, and weak driving, just above threshold i.e. $p_0$ only marginally exceeds $p_{th}$. These cases of strong and weak driving can be interpreted physically as that where the structuring nature of the instability completely dominates delocalising quantum pressure in the BEC, and that where the effects of quantum pressure are significant, respectively. In both cases we restrict the values of $b_0$, $\Delta$ etc. such that $\chi_0 n \ll 1$, for consistency with the assumption made when deriving Eq.~(\ref{eqn:GPE}) from Eq.~(\ref{eqn:sch}\ref{eqn:B}).
Fig.~\ref{fig:intens_and_dens_BEC_strong} shows an example of self-structuring displayed by the BEC-SMF model, Eq.~(\ref{eqn:sch})-(\ref{eqn:B}), in the case where the system is driven strongly, far above the instability threshold i.e. $p_0 \gg p_{th}$. The system spontaneously develops a modulated optical intensity and modulated density with a spatial period of $\Lambda_c$. The spatio-temporal distribution of the BEC density and optical intensity develop intricate "chevron" structures similar to those observed in \cite{Plestid2018} produced by a "quantum Jeans instability" \cite{Chavanis2011_2}.

\begin{figure}
\begin{center}
\includegraphics[width=\columnwidth]{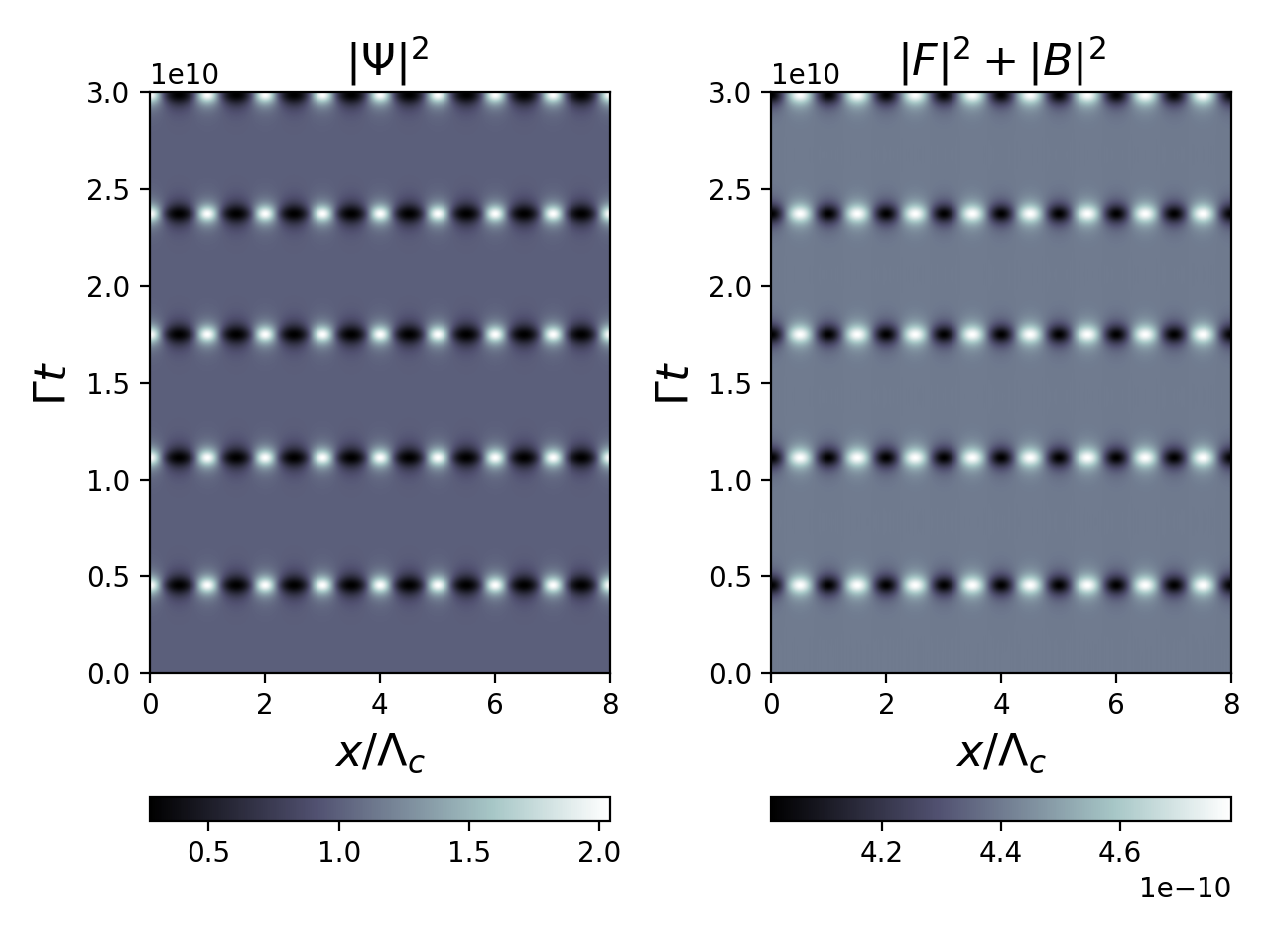}
\caption{Evolution of BEC density and optical intensity for weak driving, calculated from Eq.~(\ref{eqn:sch})-(\ref{eqn:B}). Parameters used : $b_0 = 100$, $\Delta = 500$, $p_0 = 1.1 p_{th} = 2.2 \times 10^{-10}$, $R = 1$, $\frac{\omega_r}{\Gamma}= 10^{-8}$.}
\label{fig:intens_and_dens_BEC_weak}
\end{center}
\end{figure}

Fig.~\ref{fig:intens_and_dens_BEC_weak} shows an example of self-structuring when the system is driven weakly, marginally above the instability threshold. Again, both the BEC and optical field spontaneously develop a modulation with a spatial period of $\Lambda_c$, but the evolution of the system is qualitatively different from the strongly driven case shown in Fig.~\ref{fig:intens_and_dens_BEC_strong}. In the weakly-driven case, the BEC density distribution consists of what were termed "monoclusters" in \cite{Plestid2018} and the chevrons are absent. The temporal behaviour is also different in the two cases. For weak driving, after development of the optical and BEC structures they disperse and reform regularly whereas in the strongly driven case the temporal behaviour is more complex, with a quasiperiodic sequence of dispersal and revival.

This mapping between the BEC-SMF model of Eq.~(\ref{eqn:sch})-(\ref{eqn:B}) and the quantum HMF model when $\chi_0 n \ll 1$ allows us to gain some insight into the behaviour of the BEC-SMF system. It explains the similarity in the evolution of the BEC density shown in Fig.~\ref{fig:intens_and_dens_BEC_strong} with that displayed by the quantum HMF model in \cite{Plestid2018} i.e. the chevron structures. In the weakly driven regime, it allows additional insight if we assume a wavefunction of the form
\begin{equation}
\Psi(\theta, t) = c_0(t) + c_1(t) \cos(\theta)
\label{eqn:2state}
\end{equation}
i.e. representing two states, one of which, $| 0 \rangle$, is spatially uniform, and the other, $| 1 \rangle$, which is spatially periodic with spatial period $\Lambda_c$.
Using this two-state ansatz, the effective GPE equation of the quantum HMF model in Eq.~(\ref{eqn:GPE}) can be rewritten as an equation for the order parameter/ or "magnetization" , $M$ (see supplementary material):
\begin{equation}
\left( \frac{d M}{dt} \right)^2 + \frac{\epsilon^2}{2} M^4 - \omega_r^2 \left(\frac{\epsilon}{\omega_r} - 1 \right) M^2 = 0
\end{equation}
which has the solution
\begin{equation}
M(t) = \sqrt{2} \frac{\omega_r}{\epsilon} \sqrt{\frac{\epsilon}{\omega_r} - 1} \; \mbox{sech} \left[ \omega_r \sqrt{\frac{\epsilon}{\omega_r} - 1} (t-t_0) \right]
\label{eqn:M_q_analytical}
\end{equation}
where 
$
t_0 = \frac{\cosh^{-1} \left(\frac{\sqrt{2} \frac{\omega_r}{\epsilon} 
\sqrt{
\frac{\epsilon}{\omega_r}-1}
}
{M_0} \right)}{\omega_r \sqrt{
\frac{\epsilon}{\omega_r}-1} .
}
$
 and $M_0 = M(t=0)$. 

\begin{figure}
\centering
\includegraphics[width=0.8\columnwidth]{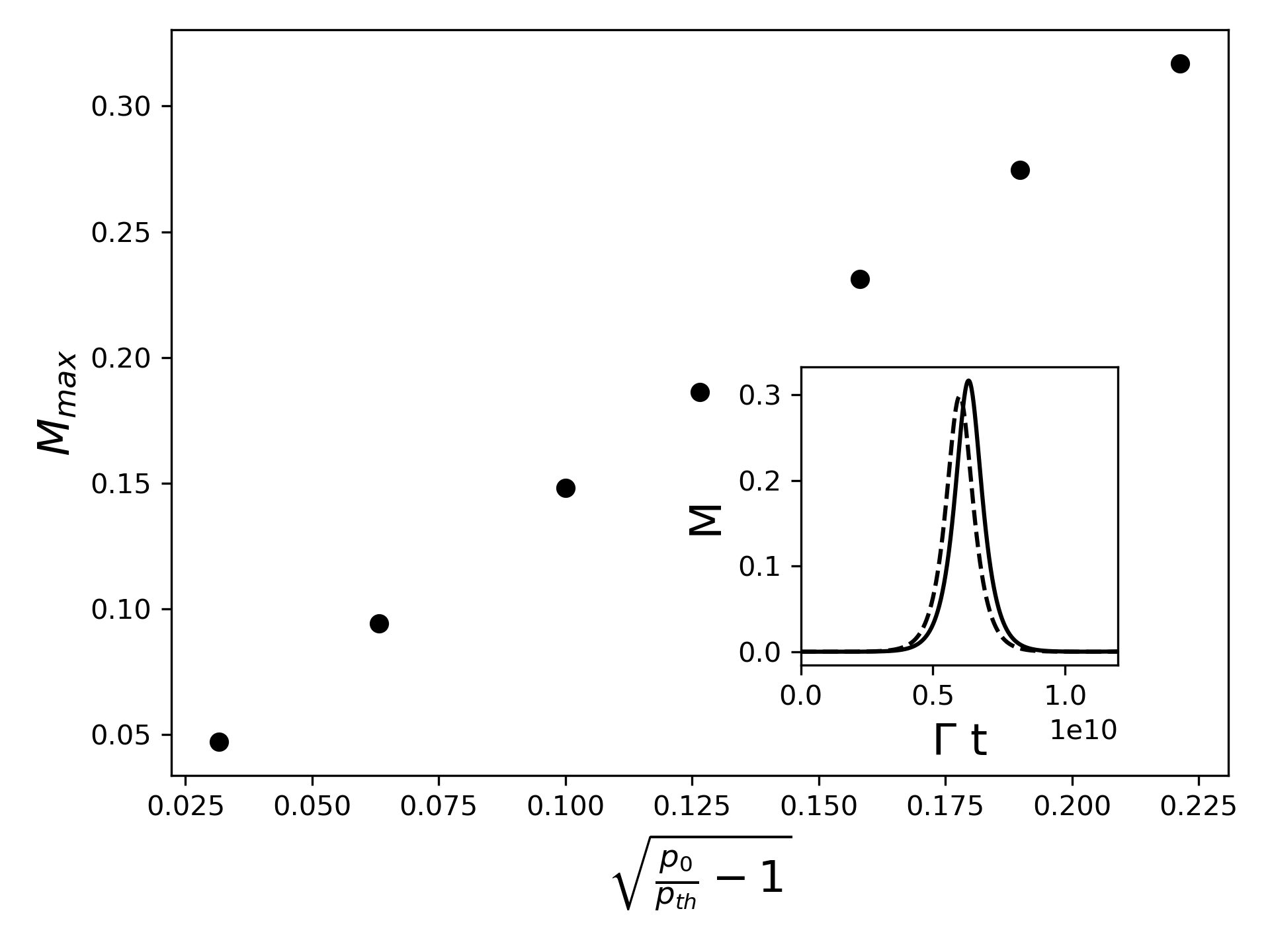}
\caption{Maximum value of $M$ as a function of $p_0$, calculated from Eq.~(\ref{eqn:sch})-(\ref{eqn:B}). All other parameters used are as for Fig.~\ref{fig:intens_and_dens_BEC_weak}. Inset shows evolution of $M$ calculated from Eq.~(\ref{eqn:M_q_analytical}) (dashed line), and from a numerical solution of Eq.~(\ref{eqn:sch})-(\ref{eqn:B})) (full line), for one period of oscillation when the system is driven weakly ($p_0 = 1.05 p_{th} = 2.098 \times 10^{-10}$). All other parameters used are as for Fig.~\ref{fig:intens_and_dens_BEC_weak}. }
\label{fig:M_vs_t_q_an_num}
\end{figure}

Fig.~\ref{fig:M_vs_t_q_an_num} (inset) shows the evolution of $M$ as calculated from Eq.~(\ref{eqn:M_q_analytical}) and from the BEC-SMF model (Eq.~(\ref{eqn:sch})-(\ref{eqn:B})), when the system is driven weakly. The analytical expression for $M$ in Eq.~(\ref{eqn:M_q_analytical}) and the numerical calculation agree well for the first period of the evolution, which in the numerical simulation then repeats periodically as in fig.~\ref{fig:intens_and_dens_BEC_weak}. The behaviour of the system in the weakly driven regime is therefore similar to that of a two-state quantum system where the BEC density (and consequently the optical intensity) oscillates spontaneously in time between a spatially uniform state and a spatially structured state. Eq.~(\ref{eqn:M_q_analytical}) predicts that the maximum value of the order parameter, $M$ scales with distance from threshold $\propto (p_0 - p_{th})^{1/2}$, similar to the mean-field Ising model. Fig.~\ref{fig:M_vs_t_q_an_num} shows that this scaling behaviour is produced by the BEC self-structuring model (Eq.~(\ref{eqn:sch})-(\ref{eqn:B})).

\begin{figure}
     \centering
     \begin{subfigure}[a]{0.45\textwidth}
         \centering
         \includegraphics[width=\columnwidth]{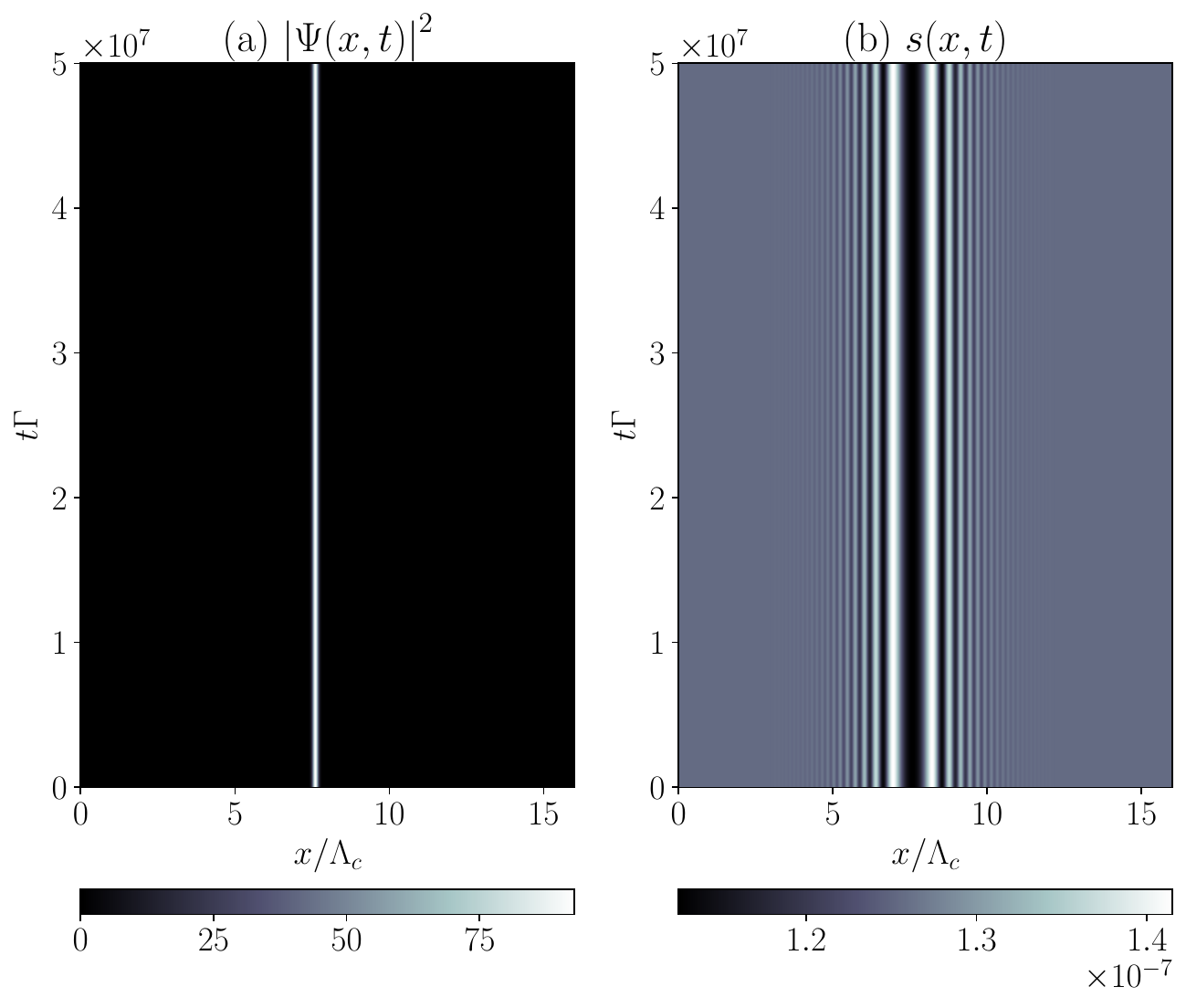}
     \end{subfigure}
     \hfill
     \begin{subfigure}[c]{0.45\textwidth}
         \centering
         \includegraphics[width=\columnwidth]{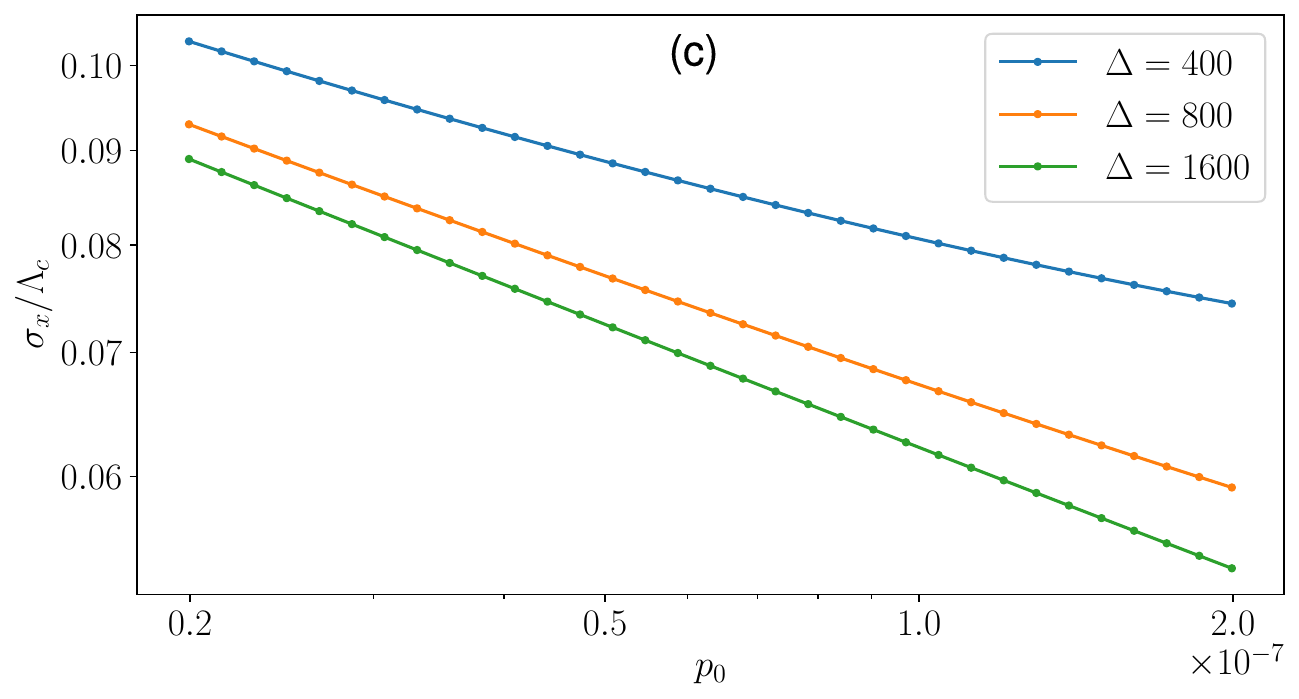}
     \end{subfigure}
     \hfill
        \caption{Evolution of (a) BEC density and (b) optical intensity distribution calculated from Eq.~(\ref{eqn:sch})-(\ref{eqn:B}) showing a stable, localised droplet. The parameters used were : $p_0=6.3 \times 10^{-8}$, $b_0=20$, $\Delta=1600$, $\omega_r/\Gamma = 1.00 \times 10^{-7}$ and $R=0.99$. The initial BEC wavefunction was Gaussian with width $\sigma_x / \Lambda_c = 0.07$. (c) Plot (on log-log scale) of stable droplet width, $\sigma_x$ vs pump intensity, $p_0$ calculated from Eq.(\ref{eqn:sch})-(\ref{eqn:B}). The parameters used were as for (a) and (b).
}
\label{fig:sigma_vs_p0}
\end{figure}

In addition to formation of global structures i.e spatially periodic patterns, it has been shown that spatially localised structures can also arise in the BEC-SMF system \cite{Zhang2018}. These structures were termed "droplets" in \cite{Zhang2018} due to the similarity with quantum droplets in dipolar BECs \cite{2016Ferrier,2016Chomaz}. An example of a stable droplet in the BEC-SMF system is shown in Fig.~\ref{fig:sigma_vs_p0} as calculated from Eq.(\ref{eqn:sch}..\ref{eqn:B}). It can be seen that a BEC of width smaller than $\Lambda_c$ maintains its shape due to its interaction with the optical field which it generates. The existence of soliton solutions for the quantum HMF model was  discovered in \cite{Plestid2019}, which showed that they are similar to strongly localised gap solitons which can exist for BECs in optical lattices \cite{Louis2003}, with the difference that in the quantum HMF model the lattice is not externally imposed, but self-generated by the BEC. Here we show that the mapping of the BEC-SMF system as described by Eq.~(\ref{eqn:sch})-(\ref{eqn:B}) to the quantum HMF model as described by Eq.~(\ref{eqn:GPE}) allows determination of the width of the droplet and its dependence on the parameters of the system e.g. pump intensity, $p_0$.

Assuming that the profile of the BEC density is Gaussian with width $\sigma_x$ i.e. $\Psi(x) \propto \exp(-\frac{ x^2}{2 \sigma_x^2})$, then the value of $\sigma_x$ which minimises the energy functional, $E(\sigma_x)$, defined as 
\begin{equation}
E(\sigma_x) = \frac{1}{2\pi} \int_0^{2 \pi} \Psi^* \left[- \omega_r \frac{\partial^2 \Psi'}{\partial \theta^2} - \epsilon \Phi(x,t) \right] \Psi \; d \theta
\end{equation}
can be shown to be (see supplementary material) 
\begin{equation}
\frac{\sigma_x}{\Lambda_c} = \frac{1}{2 \pi} \left( \frac{\omega_r}{\epsilon} \right)^{1/4} \propto (p_0 b_0 R)^{-1/4}.
\end{equation}
This is consistent with a more rigorous derivation of soliton solutions for the quantum HMF model \cite{Plestid2019}, with density profiles described by parabolic cylinder functions of characteristic width $\propto \epsilon^{-1/4}$.
The predicted dependence of droplet width, $\sigma_x$, on pump intensity, $p_0$, is confirmed in Fig.~\ref{fig:sigma_vs_p0}, where the stable droplet width is calculated from Eq.~(\ref{eqn:sch})-(\ref{eqn:B}) for different pump intensities and is plotted against $p_0$. The power-law scaling $\sigma_x \propto p_0^{-1/4}$ predicted by energy minimisation of the quantum HMF model agrees well with the results of the simulations so long as $\Delta$ is sufficiently large that condition $\chi_0 n \ll 1$ is well satisfied. This scaling behaviour shows that the profile and characteristic width of these optomechanical droplets are more closely related to those of localised gap solitons \cite{Plestid2019,Louis2003} in a self-generated lattice than to other types of localised structures e.g. non-linear Schrodinger equation solitons or quantum droplets observed in dipolar BECs \cite{2016Ferrier,2016Chomaz}.

In conclusion, we have shown that a BEC interacting with an optical field via a feedback mirror can be a realisation of the quantum HMF model. We demonstrated that the self-structuring of an initially uniform BEC displays features observed previously in the quantum HMF model: for strong driving, chevrons appear in the BEC density; for weak driving, the BEC behaves as a two-state quantum system, with the order parameter or magnetisation evolving as a series of sech pulses. The mapping to the quantum HMF model also allowed prediction the dependence of BEC droplet width on pump intensity, which agreed well with simulations of the BEC-SMF model. These results suggest that optical diffraction-mediated interaction between atoms in a BEC may be a promising candidate for experimental realisation of quantum HMF dynamics and consequently be a versatile testing ground for models of quantum systems involving long-range interactions.

\begin{acknowledgments}
We acknowledge useful discussions with G. Morigi.
\end{acknowledgments}

\end{document}


\title{Supplementary Material : Long-range interactions in a quantum gas mediated by diffracted light \\
}
\author{G.R.M. Robb, J.G.M. Walker, G.-L. Oppo \& T. A. Ackemann}
\affiliation{SUPA and Department of Physics, University of Strathclyde, Glasgow G4 0NG, Scotland, UK}

\date{\today}

\maketitle

\section*{Derivation of Magnetization Evolution for Weak Driving}
We start with the effective Gross-Pitaevskii equation (GPE) of the quantum HMF model :
\begin{equation}
i \frac{\partial \Psi}{\partial t} = 
- \omega_r \frac{\partial^2 \Psi}{\partial \theta^2} - \epsilon \Phi(\theta,t) \Psi 
\label{eqn:GPE}
\end{equation}
where $\omega_r = \frac{\hbar q_c^2}{2 m}$, $\epsilon = 2 \delta R p_0$, $\chi_0 = \frac{R p_0 b_0 \Gamma}{2}$, $\theta = q_c x$ and 
\[
\Phi(\theta,t) = \frac{1}{2 \pi} \int_{-\pi}^\pi |\Psi(\theta',t)|^2 \cos (\theta' - \theta) \; d \theta'
\] 
is a non-local potential.
The threshold value of the driving term, $\epsilon$, above which an initially homogeneous state becomes unstable is $\epsilon = \omega_r$ \cite{Plestid2018}, such that the structuring, localising effect of the non-local potential is balanced by the homogenising or delocalizing effect of the quantum pressure term. In terms of the pump intensity, $p_0$, this condition corresponds to $p_0 = p_{th}$, where $p_{th} = \frac{2 \omega_r b_0 R}{\Gamma}$.

We consider the case of a "weakly driven" instability, where the pump intensity $p_0$ only marginally exceeds the threshold value, $p_{th}$ (i.e. $\epsilon$ just exceeds $\omega_r$). In this regime, the BEC wavefunction can be written in the form 
\begin{equation}
\Psi(\theta, t) = c_0(t) + c_1(t) \cos \theta.
\label{eqn:3state}
\end{equation}
The description of the BEC is therefore entirely in terms of momentum states with momenta $0, \pm \hbar q_c$. Substituting eq.(\ref{eqn:3state}) in the quantum HMF model, eq.~(\ref{eqn:GPE}), produces 
\begin{eqnarray}
\frac{dS}{d t} &=& i \omega_r S + i \frac{\epsilon}{2} D (S + S^*) \label{eqn:dS_dt} \\
\frac{dD}{d t} &=& i \frac{\epsilon}{2} \left( S^2 - {S^*}^2\right)\label{eqn:dD_dt}
\end{eqnarray}
where $S = c_0 c_1^*$ and $D = \frac{|c_1|^2}{2} - |c_0|^2$. Writing $S = S_R + i S_I$, where $S_R$ and $S_I$ are the real and imaginary parts of $S$ respectively, then it can be shown that $S_R$, $S_I$ and $D$ satisfy the relations
\[
D^2 + 2{S_R}^2 + 2{S_I}^2 = 1
\]
and 
\[
D = \frac{\epsilon}{\omega_r} {S_R}^2 - 1
\]
for all $t$, if $S(t=0) = 0$ and $D(t=0) = - 1$ i.e. the BEC is initially spatially homogeneous. Using these relations, eq.~(\ref{eqn:dS_dt}) and (\ref{eqn:dD_dt}) can be rewritten as a single equation for $S_R$ :
\begin{equation}
\left( \frac{d S_R}{d t} \right)^2 + \frac{\epsilon^2}{2} S_R^4 - \omega_r^2 \left(\frac{\epsilon}{\omega_r} - 1 \right) S_R^2 = 0
\end{equation}
which has the solution
\[
S_R(t) = \sqrt{2} \frac{\omega_r}{\epsilon} \sqrt{\frac{\epsilon}{\omega_r} - 1} \; \mbox{sech} \left[ \omega_r \sqrt{\frac{\epsilon}{\omega_r} - 1} (t-t_0) \right]
\]
where 
\[
t_0 = \frac{\cosh^{-1} \left(\frac{\sqrt{2} \frac{\omega_r}{\epsilon} 
\sqrt{
\frac{\epsilon}{\omega_r}-1}
}
{S_R(0)} \right)}{\omega_r \sqrt{
\frac{\epsilon}{\omega_r}-1} .
}
\]
and $S_R(0) = S_R(t=0)$.

By substituting eq.(\ref{eqn:3state}) into the definition of the order parameter, magnetization i.e.
\[
M = \left| \frac{1}{2 \pi} \int_0^{2 \pi} |\Psi|^2 e^{i \theta} \; d \theta \right|,
\]
it can be shown that, $M=S_R$, so 
\[
M(t) = \sqrt{2} \frac{\omega_r}{\epsilon} \sqrt{\frac{\epsilon}{\omega_r} - 1} \; \mbox{sech} \left[ \omega_r \sqrt{\frac{\epsilon}{\omega_r} - 1} (t-t_0) \right]
\]
where 
\[
t_0 = \frac{\cosh^{-1} \left(\frac{\sqrt{2} \frac{\omega_r}{\epsilon} 
\sqrt{
\frac{\epsilon}{\omega_r}-1}
}
{M_0} \right)}{\omega_r \sqrt{
\frac{\epsilon}{\omega_r}-1} .
}
\]
and $M_0 = M(t=0)$.

\section*{Derivation of Stable Droplet Width}

The quantum Hamiltonian Mean Field (HMF) model, Eq.~(\ref{eqn:GPE}), can be written as
\begin{equation}
\label{eqn:GPE2}
i \hbar \frac{\partial \Psi(\theta,t)}{\partial t} = H \Psi
\end{equation}
where $H = \hbar \left(-\omega_r \frac{\partial^2 }{\partial \theta^2} - \epsilon \Phi \right) $.
Assuming that the BEC droplet has a spatial profile, $\Psi(\theta)$ which is a Gaussian characterised by width $\sigma$, the value of \(\sigma\) which would be expected for a stable droplet can be found by minimising the energy functional , $E(\sigma)$ corresponding to Eq.~\ref{eqn:GPE2}, defined as  
\begin{equation}
\label{eqn:defE}
E (\sigma) = \frac{1}{2 \pi} \int_{-\pi}^\pi \Psi^* H \Psi \; d \theta = \frac{\hbar}{2 \pi} \int_{-\pi}^\pi \Psi^* \left[- \omega_r \frac{\partial^2 \Psi}{\partial \theta^2} - \epsilon \Phi \Psi \right] \; d \theta .
\end{equation}

Assuming a BEC droplet with a Gaussian profile of width \(\sigma\) e.g. \[
\Psi(\theta) = C e^{-\theta^2/(2 \sigma^2)}
\] where C is a normalisation coefficient, then using the normalisation \(\frac{1}{2 \pi} \int_{-\pi}^\pi |\Psi|^2 \; d \theta = 1\), and assuming that \(\sigma \ll 2 \pi\) for simplicity, \(\int_0^{2 \pi} \approx \int_{-\infty}^\infty\) , then \[
\frac{1}{2 \pi} |C|^2 \int_{-\infty}^{\infty} e^{-\theta^2/\sigma^2} \; d \theta = \frac{1}{2 \pi} |C|^2 \sqrt{\pi} \sigma = 1 
\]
so $C = \sqrt{\frac{2}{\sigma}} \pi^{1/4}$. The assumption \(\sigma \ll 2 \pi\) implies that the droplet width is much narrower than the spatial period of the potential.

In order to evaluate \(E (\sigma) \), it is necessary to evaluate \(\frac{\partial^2 \Psi}{\partial \theta^2}\) and \(\Phi(\theta)\). From the definition of \(\Psi\) above, then \(\frac{\partial \Psi}{\partial \theta} = -\frac{1}{\sigma^2} \theta \Psi\) and \[
\frac{\partial^2 \Psi}{\partial \theta^2} = - \frac{1}{\sigma^2} \left( \Psi + \theta \frac{\partial \Psi}{\partial \theta}\right) = - \frac{1}{\sigma^2} \left( 1 - \frac{\theta^2}{\sigma^2} \right)  \Psi .
\]

Similarly, \[
\Phi(\theta) = \frac{1}{2 \pi} \int_{-\infty}^\infty |\Psi(\theta',t)|^2 \cos (\theta' - \theta) \; d \theta'
\]
so \[
\Phi(\theta) = \frac{C^2}{2 \pi} \int_{-\infty}^\infty e^{-\theta'^2 / \sigma^2} \cos (\theta' - \theta) \; d \theta' = \frac{C^2}{4 \pi} \left[ e^{-i \theta} \int_{-\infty}^\infty e^{-\theta'^2 / \sigma^2} e^{i \theta'} \; d \theta' + e^{i \theta}
\int_{-\infty}^\infty e^{-\theta'^2 / \sigma^2} e^{-i \theta'} \; d \theta'\right]
\] As the integrands are \[
e^{-\theta'^2/\sigma^2 \pm i \theta'} = \exp (-\frac{\theta'^2 \pm i \sigma^2 \theta'}{\sigma^2}) = \exp \left(-\frac{(\theta' \pm i \frac{\sigma^2}{2})^2 + \frac{\sigma^4}{4}}{\sigma^2} \right)
\] then using the identity \[
\int_{-\infty}^\infty \exp (-a(x+b)^2) \; dx = \sqrt{\frac{\pi}{a}}
\] 
,where $a$ and $b$ are constants, allows evaluation of \(\Phi\) as \[
\Phi(\theta,t) = \frac{C^2}{4 \pi} \left[ e^{-i \theta} e^{-\sigma^2/4} \sqrt{\pi} \sigma + e^{i \theta} e^{-\sigma^2/4} \sqrt{\pi} \sigma \right] = e^{-\sigma^2/4} \cos \theta.
\]

Consequently, \(E (\sigma)\) can be evaluated in Eq.~(\ref{eqn:defE}) as
\[
E(\sigma) = \frac{\hbar}{2 \pi} \int_{-\pi}^\pi\Psi^* \left[ \omega_r \frac{\partial^2 \Psi}{\partial \theta^2} - \epsilon \Phi \Psi \right] \; d \theta
\approx \frac{\hbar}{2 \pi} \int_{-\infty}^\infty  \left[ - \frac{\omega_r}{\sigma^2} \left( 1 - \frac{\theta^2}{\sigma^2} \right)  |\Psi|^2 - \epsilon e^{-\sigma^2/4} \cos \theta |\Psi|^2 \right] \; d \theta
\] so 
\begin{equation}
\label{eqn:defE2}
E (\sigma)=  \frac{\hbar \omega_r}{\sigma^2} - \frac{\hbar \omega_r}{2 \pi \sigma^4} C^2 \int_{-\infty}^\infty \theta^2 e^{-\theta^2/\sigma^2} \; d \theta - \hbar \epsilon e^{-\sigma^2/4} \frac{C^2}{2 \pi} \int_{-\infty}^\infty \cos \theta \; e^{-\theta^2/\sigma^2} \; d \theta .
\end{equation} 
 The first integral can be evaluated using the standard integral \[
\int_{-\infty}^\infty x^2 e^{-a x^2} \; dx= \frac{1}{2} \sqrt{\frac{\pi}{a^3}}
\] so that \[
\int_{-\infty}^\infty \theta^2 e^{-\theta^2/\sigma^2} \; d \theta = \frac{1}{2} \sqrt{\pi} \sigma^3 ,
\] and the second one can be evaluated as in the calculation of \(\Phi\) i.e.~ \[
C^2 \int_{-\infty}^\infty \cos \theta \; e^{-\theta^2/\sigma^2} \; d \theta =2 \pi e^{-\sigma^2/4} .
\] Substituting these into Eq.~(\ref{eqn:defE2}) produces \[
E (\sigma) = \frac{\hbar \omega_r}{\sigma^2} - \frac{\hbar \omega_r}{2 \sigma^2} - \hbar \epsilon e^{-\sigma^2/2} =  \frac{\hbar \omega_r}{2 \sigma^2} - \hbar \epsilon e^{-\sigma^2/2}
\] 

If \(\sigma \ll \sqrt{2}\), consistent with the assumption of a droplet width much narrower than the potential period, then $e^{-\sigma^2/2} \approx 1 - \frac{\sigma^2}{2} $, so 
\[
E(\sigma) \approx  \frac{\hbar \omega_r}{2 \sigma^2} - \hbar \epsilon + \frac{\hbar \epsilon \sigma^2}{2} .
\]

The condition for $E(\sigma)$ to be minimized (\(\frac{dE(\sigma)}{d \sigma} = 0\)) occurs when
$
-\frac{\omega_r}{\sigma^3} + \epsilon \sigma = 0 
$ , and consequently 
\[
\sigma = \left( \frac{\omega_r}{\epsilon} \right)^{1/4} \propto (p_0 b_0)^{-1/4} .
\] 
For consistency, as it was assumed that \(\sigma \ll \sqrt{2}\), then the above requires \(\omega_r \ll \epsilon\) i.e. a stable, narrow droplet occurs when the delocalising effect of the quantum pressure is dominated by the localising effect associated with the non-local potential.

\section*{Conditions for Neglect of Incoherent Scattering During Self-Structuring}
When light is incident on a BEC, it will scatter photons incoherently. This incoherent scattering will limit the lifetime of the BEC due to the randomly directed momentum kicks acquired by the atoms in the BEC during each photon scattering event. In order for the self-structuring instability shown in Fig.~2 and 3 to be affected negligibly by incoherent scattering, the growth rate of the self-structuring instability, $G$, should significantly exceed the photon scattering rate, $r_s$ i.e. $\frac{G}{r_s} \gg 1$.
The growth rate of the self-structuring instability, $G$, was derived in \cite{Robb2015} and can be written as 
\begin{equation}
\label{eqn:G}
G = \omega_r \sqrt{ \frac{p_0}{p_{th}}-1} .
\end{equation}
The rate at which the BEC will incoherently scatter pump photons, $r_s$ is
\[
r_s = P_e \Gamma
\]
where we have assumed two-level atoms for simplicity and $P_e$ is the probability of an atom being in its excited state. As $P_e \approx \frac{(1+R) p_0}{2}$, where $p_0$ and $R p_0$ are the saturation parameters due to the pump and backward fields respectively, then consequently, 
\begin{equation}
\label{eqn:r_s}
r_s \approx \frac{(1+R) p_0 \Gamma}{2}.
\end{equation}
From Eq.~(\ref{eqn:G}) and (\ref{eqn:r_s}), the ratio, $\frac{G}{r_s} $, can then be written as
\[
\frac{G}{r_s} = \frac{2 \omega_r}{(1+R) p_0 \Gamma} \sqrt{\frac{p_0}{p_{th}} - 1} .
\]
Writing $p_0 = \beta p_{th}$, where $\beta > 1$ to ensure that the instability is above threshold, then
\[
\frac{G}{r_s} = \frac{b_0 R}{(1+R) \beta} \sqrt{\beta - 1} \approx \frac{b_0}{2 \beta} \sqrt{\beta - 1}
\]
as $R\approx 1$ is usually assumed.
In the case of strong driving ($\beta \gg 1$), then 
$$
\frac{G}{r_s} (\rm strong) \approx \frac{b_0}{2 \sqrt{\beta}}
$$
In the example of strong driving shown in Fig.~2, $\beta = 10$, and $b_0 = 100$, so
$$
\frac{G}{r_s} (\rm strong) = 50/\sqrt{10} \approx 16
$$
In the case of weak driving ($\beta \approx 1$), then 
$$
\frac{G}{r_s} (\rm weak) \approx \frac{b_0}{2} \sqrt{\beta -1}
$$
In the example of weak driving shown in Fig.~3, $\beta = 1.1$, and $b_0 = 100$, so
$$
\frac{G}{r_s} (\rm weak) = 50 \sqrt{0.1} \approx 16
$$
The above shows that the effect of incoherent scattering of photons on the self-structuring instability shown in Figs 2 and 3 should be negligible due to $\frac{G}{r_s} \gg 1$ in both cases.
The maximum of $\frac{G}{r_s}$ occurs when $\beta = 2$, where it has the value   
\[
\frac{G}{r_s} (\rm max) = \frac{b_0 R}{2 (1+R)} \approx \frac{b_0}{4} .
\]
where $R \approx 1$ has again been assumed.